\documentclass[a4paper]{article}
\usepackage{graphicx} 
\usepackage{amsmath}
\usepackage{amssymb}
\usepackage[top=3cm, bottom=3cm, left=3cm, right=3cm]{geometry}
\usepackage{subcaption}
\usepackage{algorithm}
\usepackage[noend]{algpseudocode}
\usepackage{color}
\usepackage{hyperref}

\title{Trading with Time Series Causal Discovery:\\
An Empirical Study}
\author{Ruijie Tang \\
Business School, Imperial College London \\
\texttt{ruijie.tang23@imperial.ac.uk}}

\begin{document}

\maketitle

\begin{abstract}
This study investigates the application of causal discovery algorithms in equity markets, with a focus on their potential to build investment strategies. An investment strategy was developed based on the causal structures identified by these algorithms. The performance of the strategy is evaluated based on the profitability and effectiveness in stock markets. The results indicate that causal discovery algorithms can successfully uncover actionable causal relationships in large markets, leading to profitable investment outcomes. However, the research also identifies a critical challenge: the computational complexity and scalability of these algorithms when dealing with large datasets. This challenge presents practical limitations for their application in real-world market analysis.

\end{abstract}

\section{Introduction}
\label{sec:intro}

Numerous causal discovery algorithms have been developed for time series analysis \cite{assaad2022survey}, and some have been applied to the stock market to identify underlying driving forces \cite{huang2020causal}. However, few studies have advanced to the next step of using these driving forces in quantitative studies to predict future stock prices and develop corresponding trading strategies. This paper aims to narrow this gap by designing causality-based trading strategies and evaluating their feasibility and effectiveness.

The major contributions of this paper include the following:
\begin{enumerate}
    \item Applying time series causal discovery to real-world stock data
    \item Developing a workflow to turn causal relations into a trading strategy
    \item Backtesting the strategy on major stock markets in China and the US.
\end{enumerate}

The Python and R code developed in this research, including data pre-processing scripts, accuracy evaluation tools, causal discovery interfaces, a causal graph reformatting tool, auto-regression fitting and prediction routines, trading simulators, backtesting utilities, trading performance visualizers, and execution time profiling tools, is open-sourced\footnote{GitHub repository: \url{https://github.com/kaaaylaaa/causality-based-trading}}.

In this study, we mainly focus on three key questions:
\begin{itemize}
\item[Q1] Can we quantitatively evaluate the effectiveness of a causal discovery algorithm using stock price data?
\item[Q2] If effectiveness can be quantified, which causal discovery algorithm performs best in driving force analysis for the market?
\item[Q3] Are there practical challenges when applying causal discovery to analyze stock markets, even if the methods are theoretically sound?
\end{itemize}

The remainder of this paper is organized as follows: Section~\ref{chap:bg} provides a brief introduction to the selected causal discovery algorithms, namely tsFCI, VarLiNGAM, and TiMINo. Section~\ref{chap:methodology} outlines the detailed procedures of our workflow. Section~\ref{chap:result} presents the experimental results along with their analysis. Finally, Section~\ref{chap:conclusion} offers conclusions and recommendations for future research, and answers to the research questions.

\section{Background}
\label{chap:bg}
\begin{table}[t]
\centering
\caption{Algorithm Overview}
\begin{tabular}{llll}
\hline
Algorithm & Reference & Type             & Adaptation              \\
\hline
tsFCI     &  \cite{entner2010causal}         & Constraint-Based & Causal graph reformatting \\
VarLiNGAM &   \cite{hyvarinen2010estimation}        & Noise-Based      & Causal graph reformatting \\
TiMINo    &   \cite{peters2013causal}        & Noise-Based      & Edge direction determination \\
\hline
\end{tabular}
\end{table}

In the traditional Granger causal inference framework, a time series $Y$ causes time series $X$ if, given the past of all other time series, knowing the past of $Y$ helps in predicting future values of $X$ \cite{granger1980testing}. Although Granger causality is a widely used tool for causal inference in time series, it is limited to pairwise relationships and can be misleading in the presence of hidden common causes \cite{peters2017elements}. Time series causal discovery algorithms offer a more robust approach by addressing these limitations and providing insights into the complex causal structures within time series data \cite{assaad2022survey}.




A time series causal discovery algorithm identifies potential causal relationships within a multivariate time series and represents these relationships as a directed causal graph. In this study, we focus on the ``summary graph'' approach \cite{assaad2022survey}, where each variable in the time series corresponds to a single node in the causal graph. Previous research has applied time series causal discovery techniques to uncover driving forces \cite{huang2020causal, huang2017behind}. We aim to extend this approach by predicting future stock prices based on these driving forces. With these predictions, it becomes possible to manage an investment portfolio with various trading strategies.

Despite the numerous causal discovery approaches for time series, none excel in all situations due to the variety in data distributions. In this study, we focus on three algorithms: tsFCI, VarLiNGAM, and TiMINo, as previous research suggests they are superior to other approaches when dealing with various data structures \cite{assaad2022survey}.

\begin{itemize}
    \item tsFCI (time series Fast Causal Inference) \cite{entner2010causal} is based on the FCI algorithm \cite{spirtes2001causation}. The FCI algorithm first constructs an undirected full causal graph, removes edges using independence tests, and then orients the remaining edges according to a series of rules. This method allows for the handling of non-temporal data with hidden confounders. An extension of FCI for time series is tsFCI. Suppose we have a multivariate time series ${\mathit{X}}$ with $N$ observed variables and $T$ time points. The time lag is defined as $\tau$, and the system may contain potential hidden variables. tsFCI starts by expanding ${\mathit{X}}$ to $(T-\tau)$ rows and $(\tau+1)N$ columns using the sliding window approach. Then, the FCI algorithm is applied to find the causal structure by treating each component of ${\mathit{X}}$ as an individual random variable.
    
    \item VarLiNGAM (Vector Autoregressive Linear Non-Gaussian Acyclic Model) \cite{hyvarinen2010estimation} builds upon the LiNGAM algorithm \cite{shimizu2006linear}. The LiNGAM algorithm assumes that each variable is a linear function of its causes plus an error term $e_{i}$, which follows a non-Gaussian distribution with non-zero variance. Additionally, the error terms $e_{i}$ are independent across variables, and the system is assumed to contain no latent confounders. LiNGAM begins with a structural equation model (SEM) of the form ${\mathit{X}}=\mathbf{B}{\mathit{X}}+\mathbf{e}$ where only non-temporal data is allowed, and $\mathbf{B}$ can be permuted to a strictly lower triangular matrix with zeros on the diagonal. Solving for ${\mathit{X}}$ yields ${\mathit{X}}=\mathbf{Ae}$ where $\mathbf{A}=(\mathbf{I}-\mathbf{B})^{-1}$. However, this estimation needs proper permutation and scaling before it can be used to derive $\mathbf{B}$. Finally, the estimated $\mathbf{B}$ is permuted to strict lower triangularity, which contains the causal order. VarLiNGAM extract the causal structure by computing $\mathbf{B}_{i}$ in ${\mathit{X}}(t)=\sum_{i=0}^{\tau}\mathbf{B}_{i}{\mathit{X}}(t-i)+\mathbf{e}(t)$ using LiNGAM.
    
    \item TiMINo (Time-varying Interactions Model for Nonlinear Observations) is another SEM-based causal discovery algorithm, which considers both nonlinear and instantaneous effects \cite{peters2013causal}. A time series ${\mathit{X}}$ with absolutely continuous finite dimensional distributions w.r.t a product measure satisfies a TiMINo if $\exists \tau>0$ and $\forall X \in V$, there are sets $\mathbf{PA}_{0}^{X}\subseteq V\backslash \{X\}$, $\mathbf{PA}_{i}^{X}\subseteq V$ for $1\le i\le \tau$, such that for all $t$, $X_{t}=f_{X}((\mathbf{PA}_{\tau}^{X})_{t-\tau},\dots,(\mathbf{PA}_{1}^{X})_{t-1},(\mathbf{PA}_{0}^{X})_{t},\xi_{t}^{X})$ where $V$ is the set of variables in ${\mathit{X}}$, and $\mathbf{PA}^{X}$ represents the parent nodes of $X$ in the causal graph. $\xi^{X}_{t}$ are required to be jointly independent over both $i$ and $t$, and \textit{i.i.d.} in $t$ for each $X$. TiMINo aims to compute $\mathbf{PA}_{i}^{X}$ for each node $X$ based on the definition aforementioned.
\end{itemize}

\section{Proposed Approach}
\label{chap:methodology}

Given a target set of stocks, this empirical study involves work in three aspects: (i) data collection and processing; (ii) causal discovery, predictions and trading; and (iii) backtesting.

\subsection{Data Preparation}
\begin{table}[t]
\centering
\caption{Data Overview}
\label{data overview}
\begin{tabular}{llll}
\hline
Dataset                                & Pelosi      & CSI300      & SP500       \\
\hline
Market                                & US          & US          & China       \\
No. stocks                            & 12          & 98          & 446         \\
Start date                            & 30-Jul-2019 & 1-Sep-2009  & 1-Sep-2009  \\
End date                              & 30-Jul-2024 & 31-Dec-2019 & 31-Dec-2019 \\
No. trading days                      & 1259        & 2513        & 2604        \\
No. trading days for causal discovery & 1007        & 2010        & 2083        \\
No. trading days for backtesting      & 252         & 503         & 521    \\
\hline
\end{tabular}
\label{tab:data}
\end{table}


We selected China and the United States as two major stock markets to evaluate the effectiveness of the causal discovery algorithms. The data, sourced from \cite{li2022detecting}, covers a period of 10 years (2009.09.01 - 2019.12.31) and includes prices for SP500 and CSI300 component stocks. Another dataset that captured our interest, despite its relatively small size, holds significant potential for profitability. This dataset pertains to the stock portfolio of Nancy Pelosi, who is renowned for her astute investment insights. Notably, an ETF named `NANC' tracks her stock trades. We collected stock prices from her current portfolio, comprising 12 stocks, over the past five years from Yahoo Finance \footnote{Yahoo Finance: \url{https://finance.yahoo.com/}}. 

We performed data imputation due to the substantial amount of missing values in the datasets. The imputation process involved two steps. First, linear interpolation was applied across all stocks to estimate and fill gaps where missing values occurred between known values. Second, any stock that still contained missing values after interpolation was removed from the dataset. As a result, the cleaned SP500 dataset comprises 446 stocks, while the cleaned CSI300 dataset comprises 98 stocks. The Pelosi dataset contains no missing values, so no stocks were removed. An overview of our data is in Table~\ref{data overview}.

\subsection{Causal Discovery, Prediction and Trading}


The three causal discovery algorithms discussed in Section~\ref{chap:bg}, namely tsFCI, VarLiNGAM and TiMINo, are employed to extract the causal structure from each dataset. Our primary focus is on identifying the causal relationships among the time series rather than the exact causal time lags. Therefore, the causal graphs generated by tsFCI and VarLiNGAM are compressed by removing the time lag attributes for subsequent analysis.




\begin{figure}[t]
    \centering
    \includegraphics[width=\linewidth]{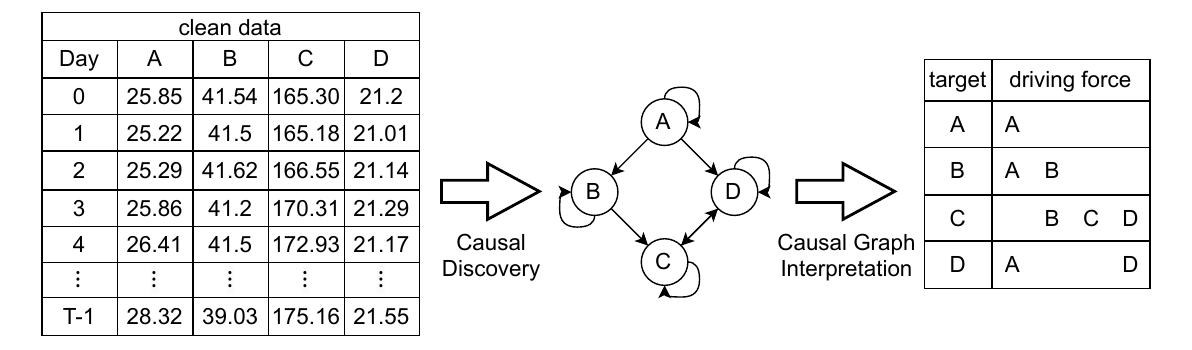}
    \caption{Use causal discovery to find driving forces}
    \label{fig:causal_discovery}
\end{figure}

\begin{figure}[t]
    \centering
    \includegraphics[width=\linewidth]{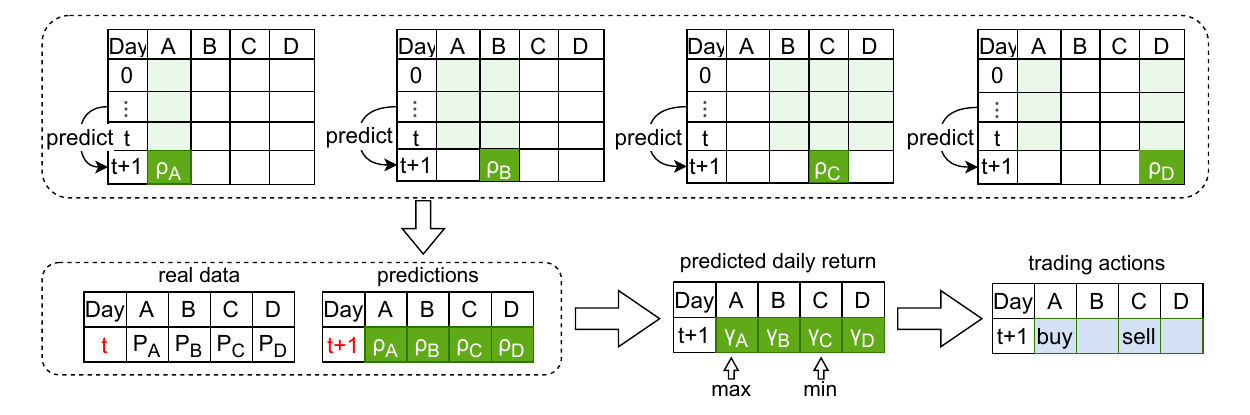}
    \caption{1-day ahead predictions and trading actions}
    \label{1day}
\end{figure}

The process of using causal discovery techniques to identify driving forces for each stock is illustrated in Figure \ref{fig:causal_discovery}. From the causal graph, we can extract the set of parent nodes of stock $X$, denoted by $\mathbf{PA}^{X}$, which can be interpreted as the driving forces of $X$ \cite{huang2020causal}. Then, similar to the multivariate time series model described by \cite{velu2020algorithmic}, we fit a predictive model to each stock using $\mathbf{PA}^{X}$ as independent variables. The model is formulated as follows:
\begin{equation}
    P^{X}_{t}=f_{X}(\mathbf{PA}^{X}_{t-1},...,\mathbf{PA}^{X}_{t-\tau})
\end{equation}
where $\tau$ is the time lag used during causal discovery. The latest time point we use for prediction is $t-1$ instead of $t$. In other words, the current value is modelled by the changes in the driving forces at previous time steps, while the present does not produce an instantaneous effect. An expanding window is employed when fitting the predictive model. The daily prediction and trading process is outlined in Figure \ref{1day}. For each stock, we extract the historical values of its driving forces, fit a linear regression model on these past prices, and then make a one-step ahead prediction. Using the predicted prices, we calculate the predicted returns for the next day as follows:
\begin{equation}
    \gamma^{X}_{t,t+1}=\frac{\rho^{X}_{t+1}-P^{X}_{t}}{P^{X}_{t}}
\end{equation}
where $\rho^{X}_{t+1}$ is the predicted price of stock $X$ for the next day, and $P^{X}_{t}$ is the actual stock price today. Finally, we aggregate the predictions of all stocks in the asset class for the next day and proceed with the corresponding trading actions.


Given price predictions, numerous studies have demonstrated that a long-short momentum strategy can generate significant profits \cite{jegadeesh1993returns, wiest2023momentumReview, moskowitz1999industries, moskowitz2012time}. This strategy operates on the premise that stocks with strong past performance are likely to continue performing well, while those with poor past performance are expected to underperform. The cross-sectional long-short momentum strategy introduced by \cite{jegadeesh1993returns} constructs a zero-cost market-neutral portfolio by buying stocks with high past returns (winners) and selling those with low past returns (losers).

Based on this approach and integrating short-term forecasting, our trading strategy employs a dollar-neutral portfolio, ensuring that the dollar amounts of long and short positions are balanced \cite{chincarini2006quantitative, asness2013value}. For simplicity, we assume a liquid market where short selling is permitted without incurring borrowing costs. However, given that our strategy operates on a daily basis, transaction costs cannot be ignored as they can significantly impact portfolio returns. Therefore, we impose a fixed daily transaction cost of 0.1\% \cite{frazzini2018trading}. The strategy is implemented as follows:



First, we select an integer $\eta$ as the number of winner stocks. Then, using the one-day-ahead predictions obtained through regression, we identify the $\eta$ stocks with the highest (lowest) predicted returns, classifying them as winners (losers). Finally, at the end of the current trading day, we buy the winners and sell the losers. This approach allows us to construct a zero-cost, dollar-neutral portfolio with equally weighted stocks on both the long and short sides. All trading positions are closed out at the end of each day before the next trading action takes place.

\subsection{Back-testing}
At this stage, we use historical data to evaluate the performance of the trading strategy. The procedure follows the steps outlined in the previous two sections. To avoid look-ahead bias, we use the earliest 80\% of the dataset as the training set and reserve the remaining 20\% as the test set, as detailed in Table~\ref{tab:data}. In each trading day, we run the trading strategy and compute the outcome according to the real data. In particular, the realized portfolio daily return is calculated as follows:
\begin{equation}
r^{p}_{t,t+1}=\frac{r^{L_{1}}_{t,t+1}+\dots+r^{L_{\eta}}_{t,t+1}}{\eta}-\frac{r^{S_{1}}_{t,t+1}+\dots+r^{S_{\eta}}_{t,t+1}}{\eta}- C
\end{equation}
where $C$ is the transaction cost; $r^{L},r^{S}$ represent the realized returns of the long and short positions, respectively. We use the expanding window method to continue with one-step ahead forecasts until the end of test period. Finally, the annualized portfolio return is calculated as follows \cite{chincarini2006quantitative}:
\begin{equation}
    r^{p}_{annual}=(1+r^{p}_{T_{test}})^{\frac{D}{T_{test}}}-1
\end{equation}
where $r^{p}_{T_{test}}$ represents the cumulative return over the test period, $T_{test}$ is the length of the test period, and $D=252$ is the number of trading days in a year.


\section{Evaluation}
\label{chap:result}
\subsection{Raw Results}

Before applying the causal discovery algorithms to analyze the stock market, we validate the correctness of our setup using datasets with ground truth causal graphs, achieving the similar model accuracy scores to those reported in \cite{assaad2022survey}.

\begin{table}[h]
\centering
\caption{Raw Result Placement}
\label{tab:raw result}
\begin{tabular}{llll}
\hline
                        & Pelosi & CSI300               & SP500                 \\
                        \hline
tsFCI (lags 1--2)       & Figure~\ref{Pelosi portfolio performance using VarLiNGAM, tsFCI and TiMINo} & Cannot finish in 24h & Cannot finish in 24h  \\
tsFCI (lags 3--6)       & Appendix & Cannot finish in 24h & Cannot finish in 24h  \\
TiMINo   (lags 1--2)    & Figure~\ref{Pelosi portfolio performance using VarLiNGAM, tsFCI and TiMINo} & Cannot finish in 24h & Cannot finish in 24h  \\
TiMINo   (lags 3--6)    & Appendix & Cannot finish in 24h & Cannot finish in 24h  \\
VarLiNGAM   (lags 1--2) & Figure~\ref{Pelosi portfolio performance using VarLiNGAM, tsFCI and TiMINo} & Figure~\ref{SP500 and CSI300 portfolio performance using VarLiNGAM}               & Figure~\ref{SP500 and CSI300 portfolio performance using VarLiNGAM}                \\
VarLiNGAM   (lag 3) & Appendix & Figure~\ref{SP500 and CSI300 portfolio performance using VarLiNGAM}               & Appendix                \\
VarLiNGAM   (lag 4) & Appendix & Figure~\ref{SP500 and CSI300 portfolio performance using VarLiNGAM}               & Memory limit exceeded                \\
VarLiNGAM (lags 5--6)   & Appendix & Appendix               & Memory limit exceeded\\
\hline
\end{tabular}
\end{table}

\begin{figure}[H]
    \centering
    \begin{subfigure}[b]{0.49\textwidth}
        \centering
        \includegraphics[width=\textwidth]{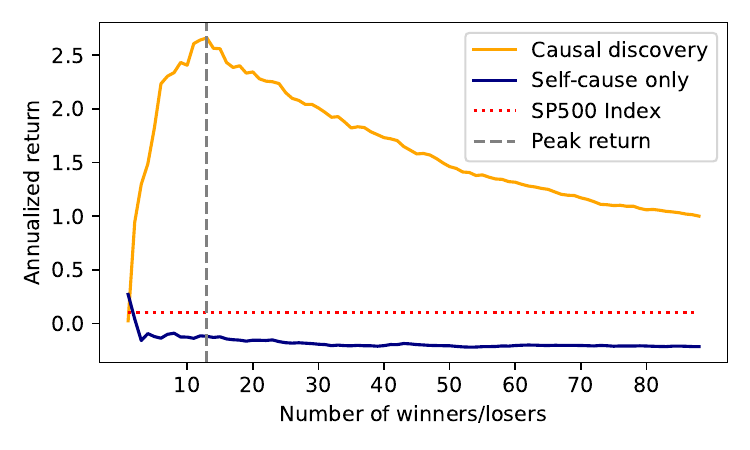}
        \caption{VarLiNGAM, SP500, lag=1}
    \end{subfigure}    
    \begin{subfigure}[b]{0.49\textwidth}
        \centering
        \includegraphics[width=\textwidth]{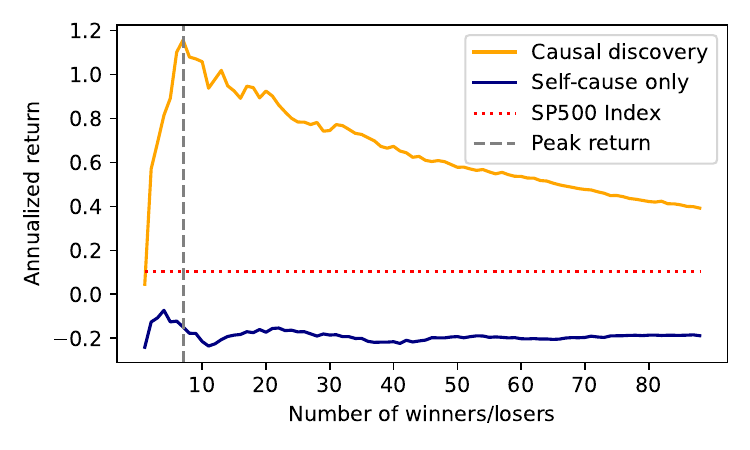}
        \caption{VarLiNGAM, SP500, lag=2}
    \end{subfigure}
    \begin{subfigure}[b]{0.49\textwidth}
        \centering
        \includegraphics[width=\textwidth]{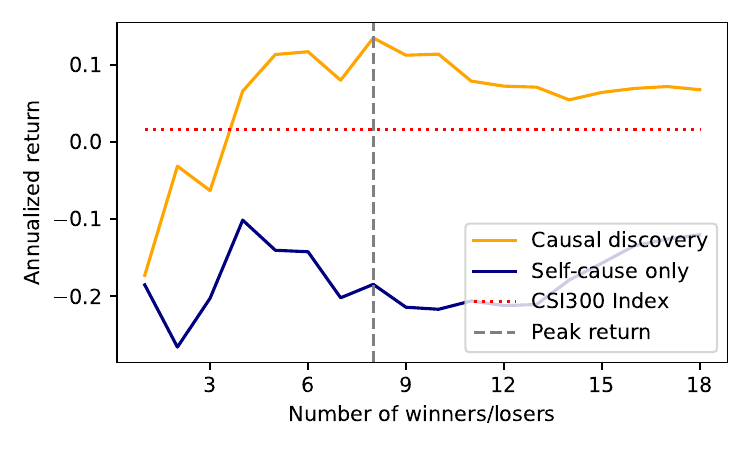}
        \caption{VarLiNGAM, CSI300, lag=1}
    \end{subfigure}    
    \begin{subfigure}[b]{0.49\textwidth}
        \centering
        \includegraphics[width=\textwidth]{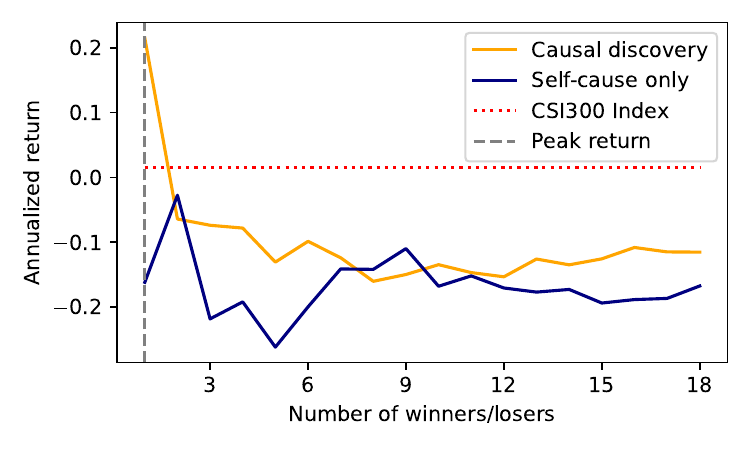}
        \caption{VarLiNGAM, CSI300, lag=2}
    \end{subfigure}
    \begin{subfigure}[b]{0.49\textwidth}
        \centering
        \includegraphics[width=\textwidth]{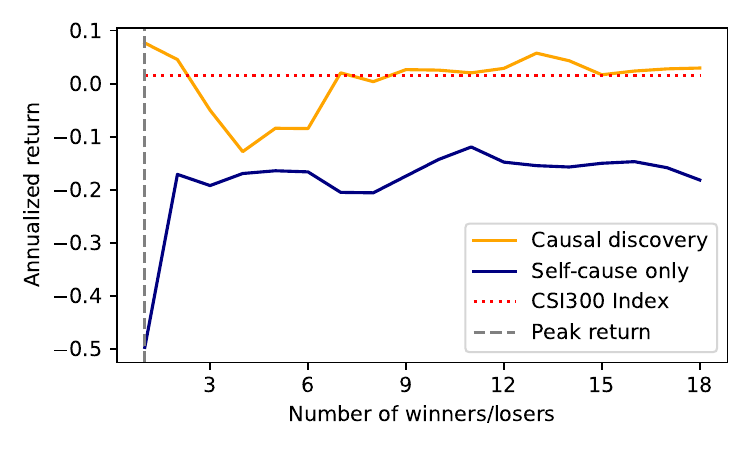}
        \caption{VarLiNGAM, CSI300, lag=3}
    \end{subfigure}
    \begin{subfigure}[b]{0.49\textwidth}
        \centering
        \includegraphics[width=\textwidth]{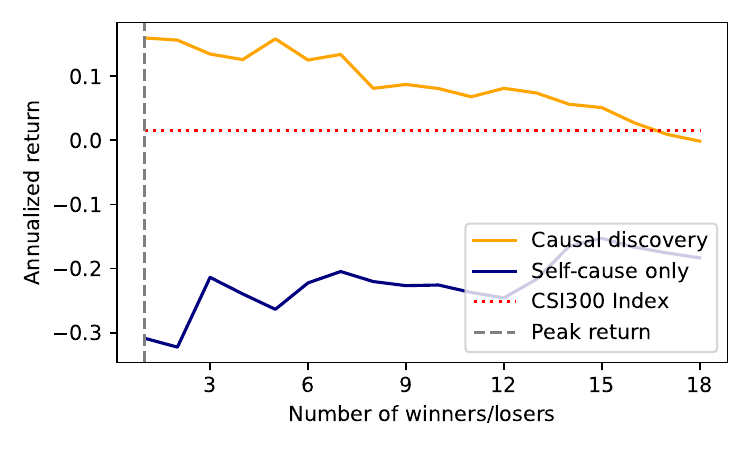}
        \caption{VarLiNGAM, CSI300, lag=4}
    \end{subfigure}

    \caption{SP500 \& CSI300 portfolio performance using VarLiNGAM}
    (Discussion available in Section~\ref{sec:discussion})
    \label{SP500 and CSI300 portfolio performance using VarLiNGAM}
\end{figure}

We record the performance for a test case if the algorithm is able to finish execution within 24 hours on an Apple MacBook Air computer with Apple M2 chips and 16GB of memory. Table~\ref{tab:raw result} provides a summary of our tests. Both tsFCI and TiMINo were unable to process the CSI300 and SP500 datasets due to excessive execution time. For the successful runs, we aim to test lags from 1 to 6. However, while processing the SP500 dataset with VarLiNGAM using time lags from 4 to 6, we encountered a memory limit issue. The results for VarLiNGAM on the CSI300 (lags 1 to 4) and SP500 (lags 1 and 2) datasets are shown in Figure \ref{SP500 and CSI300 portfolio performance using VarLiNGAM}. The results for all three algorithms with the first two lags for the Pelosi dataset are shown in Figure \ref{Pelosi portfolio performance using VarLiNGAM, tsFCI and TiMINo}. The remaining graphs are provided in the Appendix.


For ease of comparison, all returns were standardized to annualized returns. Each asset class was benchmarked against its respective index or ETF to evaluate portfolio performance. To assess the validity and necessity of causal discovery, a control portfolio was constructed using predictions based solely on self-causality. 

\begin{figure}[H]
    \centering
    
    \begin{subfigure}[b]{0.49\textwidth}
        \centering
        \includegraphics[width=\textwidth]{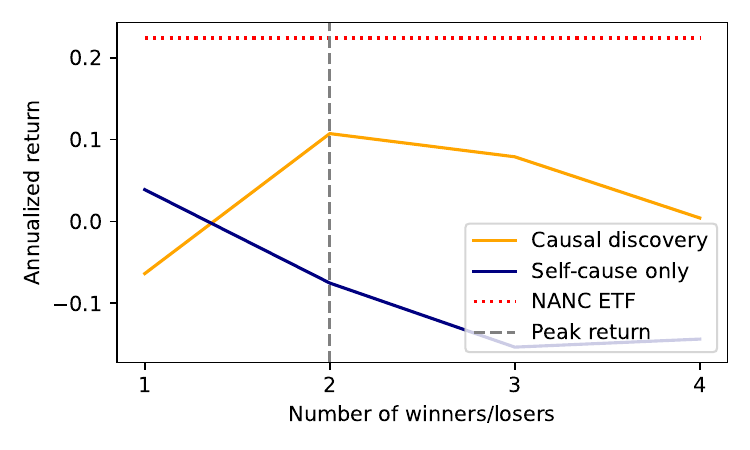}
        \caption{Pelosi, VarLiNGAM, lag=1}
    \end{subfigure}
    \begin{subfigure}[b]{0.49\textwidth}
        \centering
        \includegraphics[width=\textwidth]{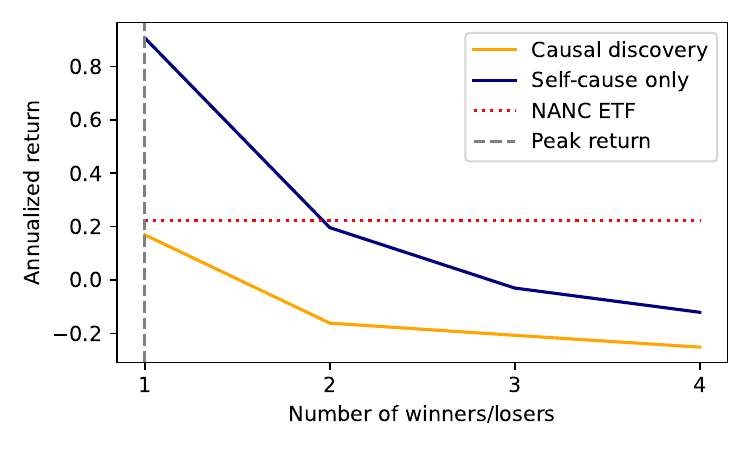}
        \caption{Pelosi, VarLiNGAM, lag=2}
    \end{subfigure}
    \begin{subfigure}[b]{0.49\textwidth}
        \centering
        \includegraphics[width=\textwidth]{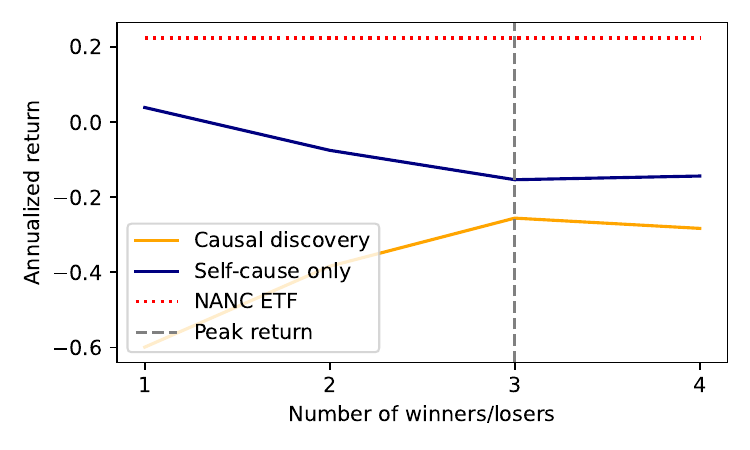}
        \caption{Pelosi, tsFCI, lag=1}
    \end{subfigure}
    \begin{subfigure}[b]{0.49\textwidth}
        \centering
        \includegraphics[width=\textwidth]{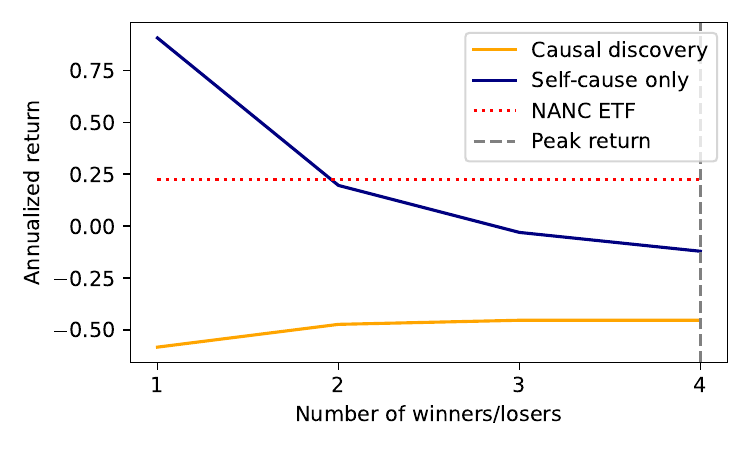}
        \caption{Pelosi, tsFCI, lag=2}
    \end{subfigure}
    \begin{subfigure}[b]{0.49\textwidth}
        \centering
        \includegraphics[width=\textwidth]{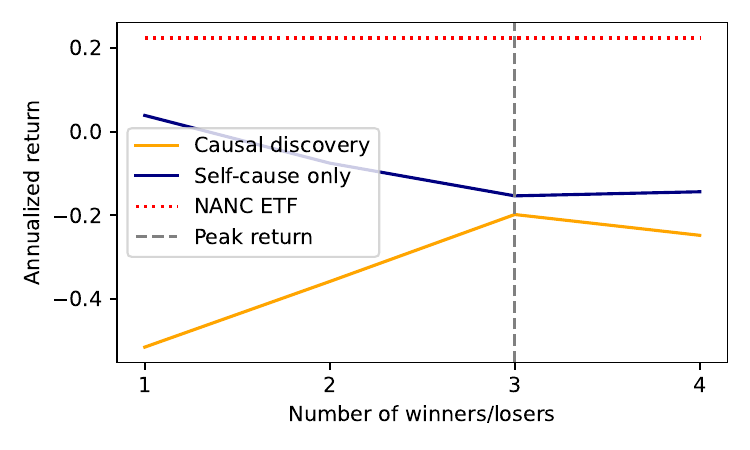}
        \caption{Pelosi, TiMINo, lag=1}
    \end{subfigure}
    \begin{subfigure}[b]{0.49\textwidth}
        \centering
        \includegraphics[width=\textwidth]{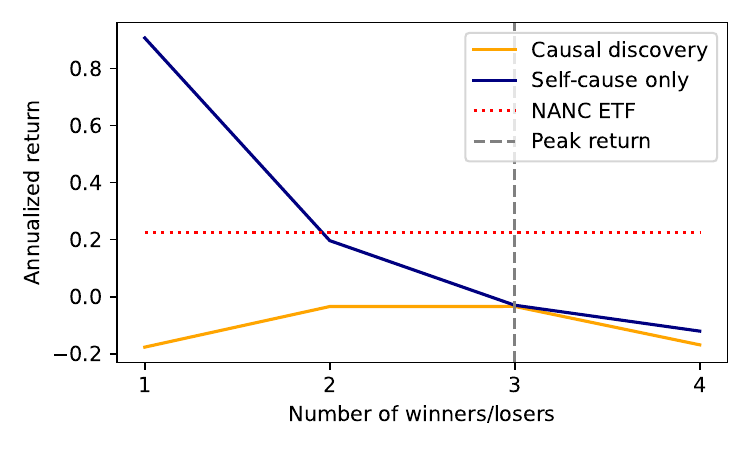}
        \caption{Pelosi, TiMINo, lag=2}
    \end{subfigure}

    \caption{Pelosi portfolio performance using VarLiNGAM, tsFCI and TiMINo}
    (Discussion available in Section~\ref{sec:discussion})
    \label{Pelosi portfolio performance using VarLiNGAM, tsFCI and TiMINo}
\end{figure}

\subsection{Discussion}
\label{sec:discussion}

Our results indicate that incorporating a causal discovery phase generally improves portfolio returns, especially with larger datasets. Among the three algorithms, VarLiNGAM is the most effective, handling large datasets well and generating the highest returns on all datasets. Major observations we obtain from the results include the following:
\begin{enumerate}
    \item \textbf{Causal discovery proves more effective than the self-cause-only setting.} The portfolio supported by causal discovery significantly outperforms the self-cause-only portfolio. Conversely, the self-cause-only portfolio performs poorly, providing worse performance even than the baseline index returns. In other words, relying solely on the stock's own historical data for price prediction is sub-optimal. This observation is particularly clear in Figure \ref{SP500 and CSI300 portfolio performance using VarLiNGAM} with a large number of stocks.
    \item \textbf{Trading with causal discovery tends to be more profitable in large markets than small markets.} Comparing Figure \ref{SP500 and CSI300 portfolio performance using VarLiNGAM} and \ref{Pelosi portfolio performance using VarLiNGAM, tsFCI and TiMINo}, we find that the trading strategy based on causal discovery becomes more effective with the markets with more choices. A small dataset like Pelosi hinders the causal discovery process by including too few stocks, likely excluding the true driving forces from the analysis.
    \item \textbf{A smaller time lag yields higher returns.} The results from all three datasets indicate that a time lag of 1 or 2 yields the most favorable portfolio returns, with portfolio returns generally decreasing as the time lag increases. This outcome may be attributed to the increase in the number of independent variables considered in the predictive model as the time lag grows. Such an increase can lead to potential issues of overfitting or expose the limitations of the simple linear regression model chosen.
    \item \textbf{The optimal number of winners ranges from 1\% to 6\% of the total stocks in the dataset.} In the SP500 market, the portfolio returns display a humped pattern across various time lags. As the number of winners increases, the portfolio return initially rises, reaching a peak due to diversification, which enhances returns by reducing risk. However, beyond this point, further increasing the number of winners dilutes returns, as the portfolio becomes overly diversified with too many different stocks. In contrast, analyzing the number of winners for the Pelosi dataset seems less relevant, given the poor performance of the causal discovery portfolio. 
    \item \textbf{VarLiNGAM achieves the highest returns among the three algorithms.} VarLiNGAM is the only method that can finish computation within 24 hours for SP500 and CSI300. Also, in the Pelosi dataset where all three methods provide valid results, VarLiNGAM still has minor advantages over tsFCI and TiMINo, especially when the lag is 1.
\end{enumerate}

\section{Conclusion}
\label{chap:conclusion}
In conclusion, among the three tested causal discovery algorithms, VarLiNGAM is the only feasible option for analyzing the stock market with 400--500 stocks, while tsFCI and TiMINo are constrained to processing much smaller datasets. For a large collection of stocks like SP500 or CSI300 constituents, causal discovery is helpful for improving prediction accuracy and enhancing trading performance. The overall portfolio performance in the SP500 constituents surpasses that of CSI300, which is expected given the bullish overall performance of the SP500 Index. On the other hand, causal discovery proves ineffective when applied to smaller asset classes like Pelosi, where the true driving forces of its component stocks fall outside the analysis scope. Regardless of the market size, VarLiNGAM still outperforms tsFCI and TiMINo in all analysed cases, making it the best algorithm among the three. In terms of trading, it is advisable to select a smaller time lag. Specifically, for our trading strategy, setting the number of winner stocks to 1\%-6\% of the size of the original asset class yields the best returns. Future work of this study includes the following directions:
\begin{itemize}
\item[F1] We explored the feasibility of developing trading strategies based on causal discovery techniques. Among the three algorithms examined, only VarLiNGAM demonstrated the capability to efficiently process large datasets. Consequently, a direction for future research is to improve the computationally efficiency and scalability of time series causal discovery techniques with respect to the data size.
\item[F2] Once the causal structure is established, predictions can be made based on the identified causal relationships. This study uses a simple linear regression model for the preditions, which is known to have limited predictive power. Future research should consider more advanced predictive models like Long short-term memory (LSTM) models \cite{fischer2018deep, ballings2015evaluating}.
\end{itemize}

We finish this paper by answering the research questions specified in Section~\ref{sec:intro}:
\begin{itemize}
\item [A1] It is feasible to quantitatively analyze the effectiveness of a causal discovery algorithm using stock price data by constructing a trading portfolio based on the causal graph and then tracking the portfolio's returns over time.
\item [A2] VarLiNGAM emerged as the most effective algorithm among the three selected methods for two key reasons. First, it is capable of processing large datasets, which the other algorithms cannot handle. Second, the trading portfolio constructed using VarLiNGAM outperforms those based on the other two algorithms, generating the highest returns on the smaller Pelosi dataset. Further testing of additional algorithms is required to confirm these findings.\\
\item [A3] The primary challenge in applying causal discovery to the stock market lies in computational limitations. Among the three algorithms considered, only VarLiNGAM is capable of analyzing large datasets such as CSI300 and SP500 within 24 hours, while tsFCI and TiMINo struggle to process such extensive equity markets. Similar computational bottlenecks have been recognized and addressed in non-temporal causal discovery research \cite{guo2022accelerating, hagedorn2022gpu}. However, faster and more scalable time series causal discovery techniques should be developed to handle data from large markets within limited time and resources, as we discussed in Future Work F1.
\end{itemize}

\bibliographystyle{plain}
\bibliography{references}

\appendix
\section*{Appendix}
The following graphs are discussed in Section~\ref{sec:discussion}.
\begin{figure}[htbp]
    \centering
    \includegraphics[width=0.49\linewidth]{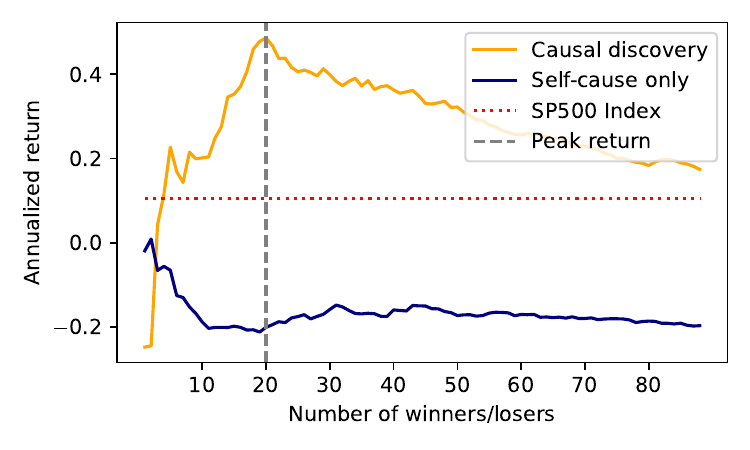}
    \caption{SP500 portfolio performance using VarLiNGAM, lag=3}
\end{figure}

\begin{figure}[htbp]
    \centering
    \begin{subfigure}[b]{0.49\textwidth}
        \centering
        \includegraphics[width=\textwidth]{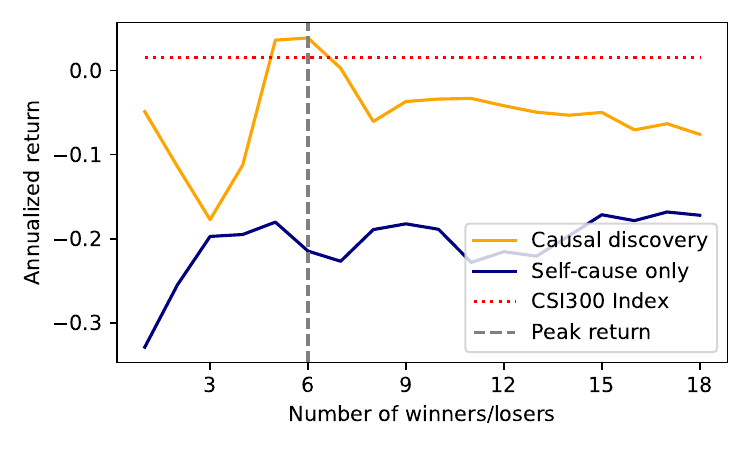}
        \caption{VarLiNGAM, CSI300, lag=5}
    \end{subfigure}
    \begin{subfigure}[b]{0.49\textwidth}
        \centering
        \includegraphics[width=\textwidth]{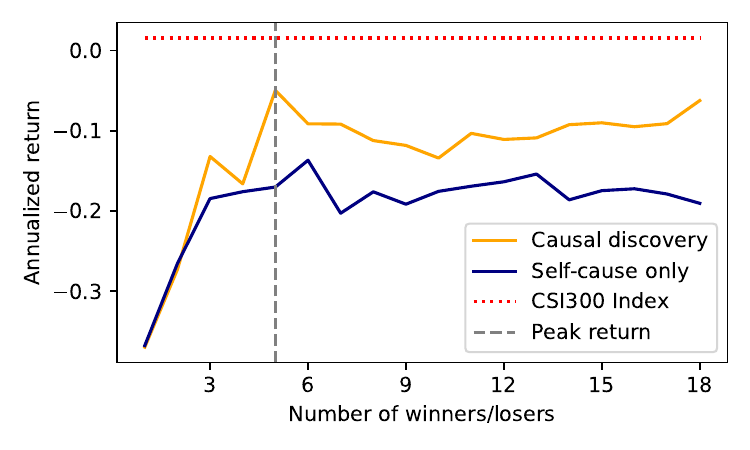}
        \caption{VarLiNGAM, CSI300, lag=6}
    \end{subfigure}

    \caption{CSI300 portfolio performance using VarLiNGAM}
\end{figure}

\begin{figure}[htbp]
    \centering
    
    \begin{subfigure}[b]{0.49\textwidth}
        \centering
        \includegraphics[width=\textwidth]{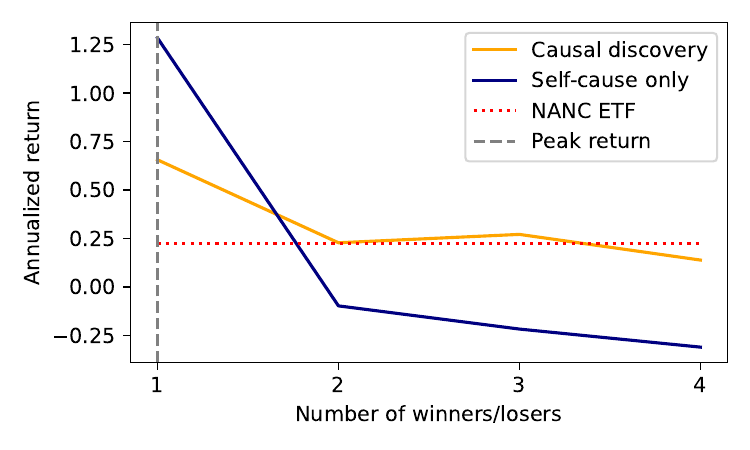}
        \caption{Pelosi, VarLiNGAM, lag=3}
    \end{subfigure}
    \begin{subfigure}[b]{0.49\textwidth}
        \centering
        \includegraphics[width=\textwidth]{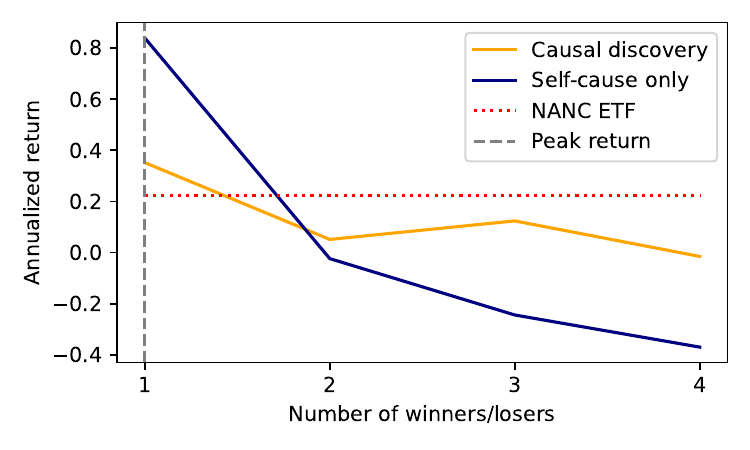}
        \caption{Pelosi, VarLiNGAM, lag=4}
    \end{subfigure}
    \begin{subfigure}[b]{0.49\textwidth}
        \centering
        \includegraphics[width=\textwidth]{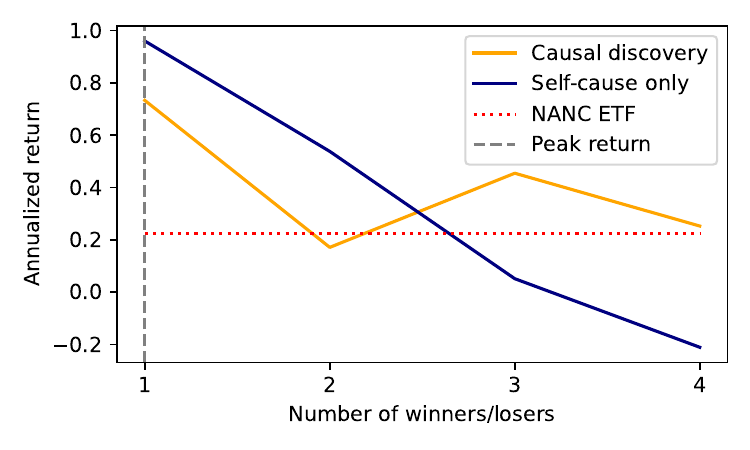}
        \caption{Pelosi, VarLiNGAM, lag=5}
    \end{subfigure}
    \begin{subfigure}[b]{0.49\textwidth}
        \centering
        \includegraphics[width=\textwidth]{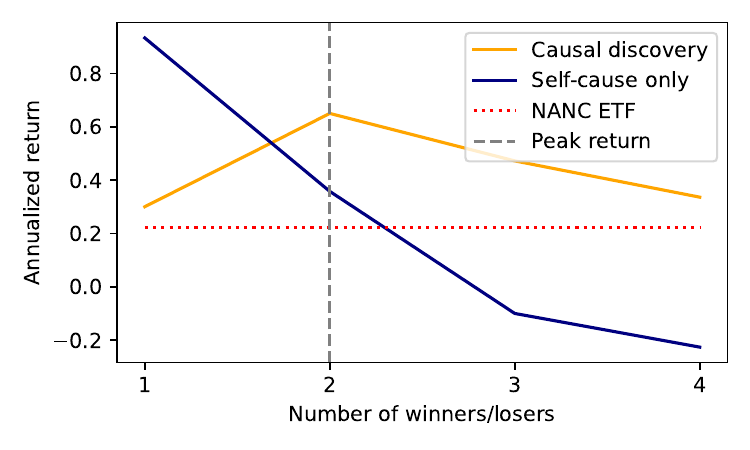}
        \caption{Pelosi, VarLiNGAM, lag=6}
    \end{subfigure}
    \caption{Pelosi portfolio performance using VarLiNGAM}
\end{figure}

\begin{figure}[htbp]
    \centering
    
    \begin{subfigure}[b]{0.49\textwidth}
        \centering
        \includegraphics[width=\textwidth]{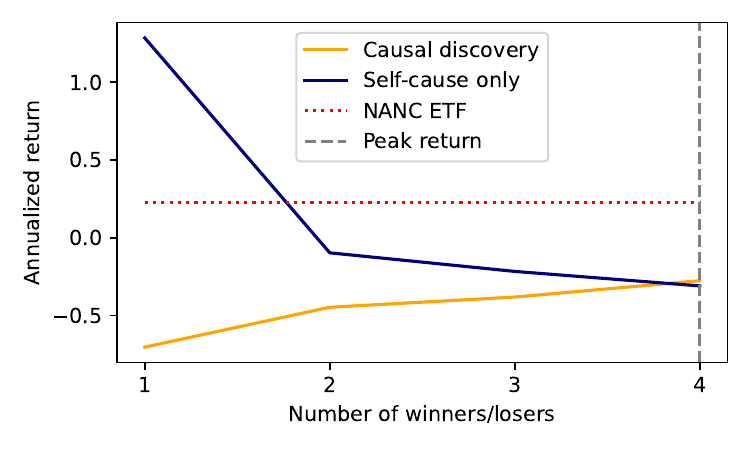}
        \caption{Pelosi, tsFCI, lag=3}
    \end{subfigure}
    \begin{subfigure}[b]{0.49\textwidth}
        \centering
        \includegraphics[width=\textwidth]{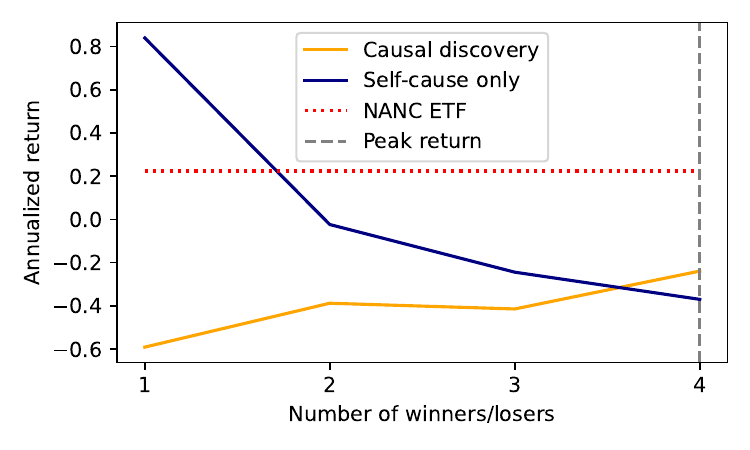}
        \caption{Pelosi, tsFCI, lag=4}
    \end{subfigure}
    \begin{subfigure}[b]{0.49\textwidth}
        \centering
        \includegraphics[width=\textwidth]{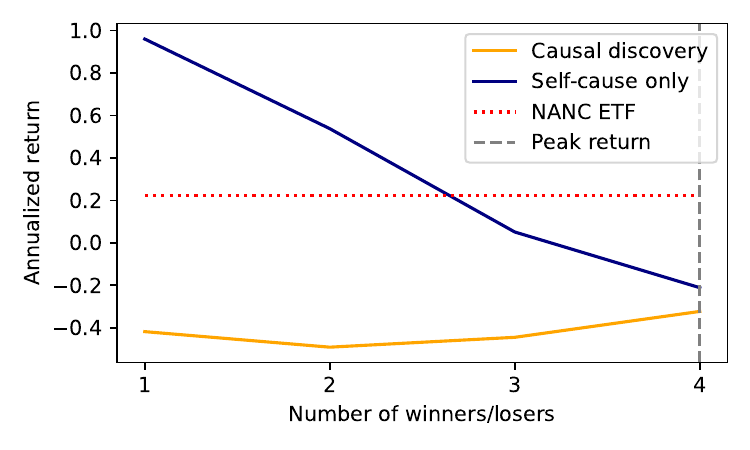}
        \caption{Pelosi, tsFCI, lag=5}
    \end{subfigure}
    \begin{subfigure}[b]{0.49\textwidth}
        \centering
        \includegraphics[width=\textwidth]{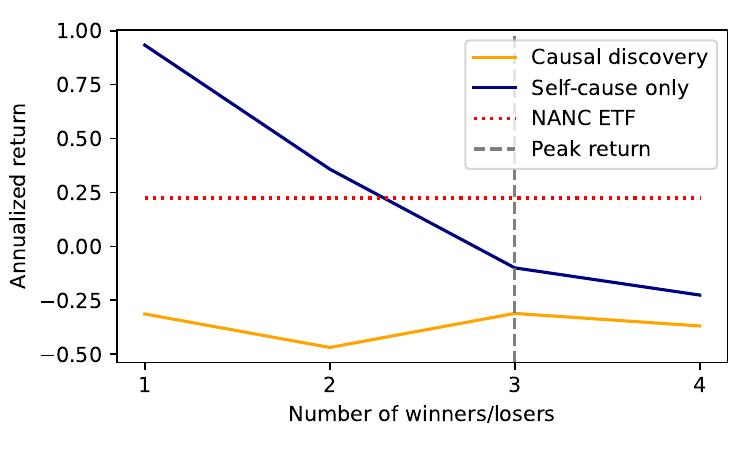}
        \caption{Pelosi, tsFCI, lag=6}
    \end{subfigure}
    \caption{Pelosi portfolio performance using tsFCI}
\end{figure}

\begin{figure}[htbp]
    \centering
    
    \begin{subfigure}[b]{0.49\textwidth}
        \centering
        \includegraphics[width=\textwidth]{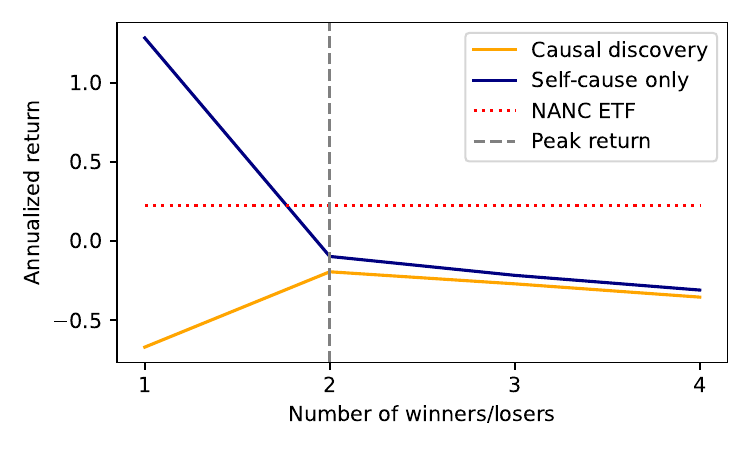}
        \caption{Pelosi, TiMINo, lag=3}
    \end{subfigure}
    \begin{subfigure}[b]{0.49\textwidth}
        \centering
        \includegraphics[width=\textwidth]{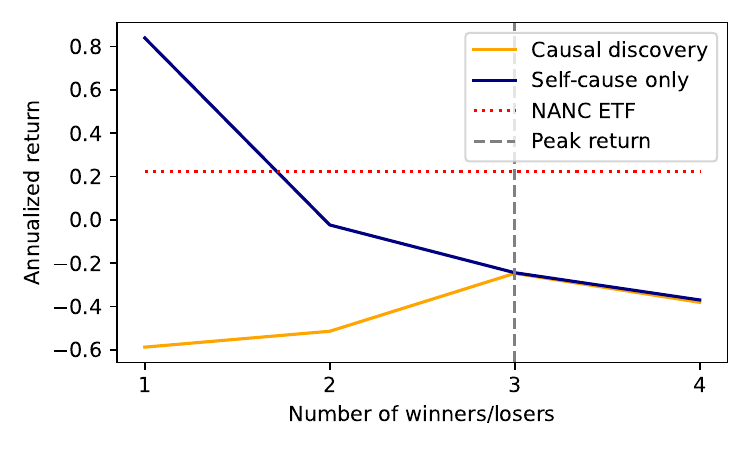}
        \caption{Pelosi, TiMINo, lag=4}
    \end{subfigure}
    \begin{subfigure}[b]{0.49\textwidth}
        \centering
        \includegraphics[width=\textwidth]{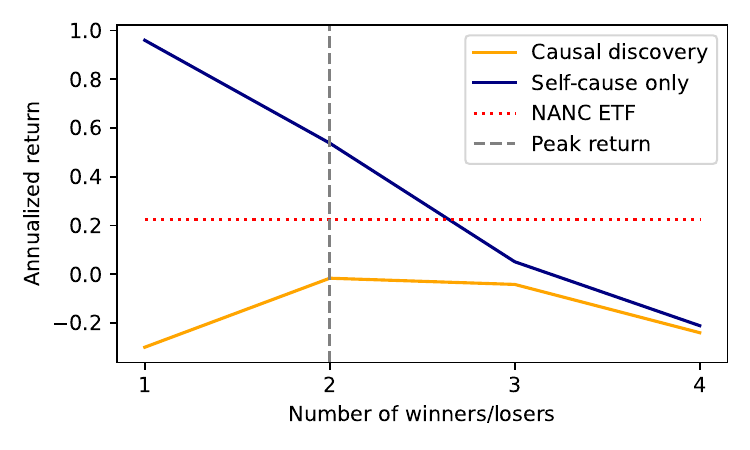}
        \caption{Pelosi, TiMINo, lag=5}
    \end{subfigure}
    \begin{subfigure}[b]{0.49\textwidth}
        \centering
        \includegraphics[width=\textwidth]{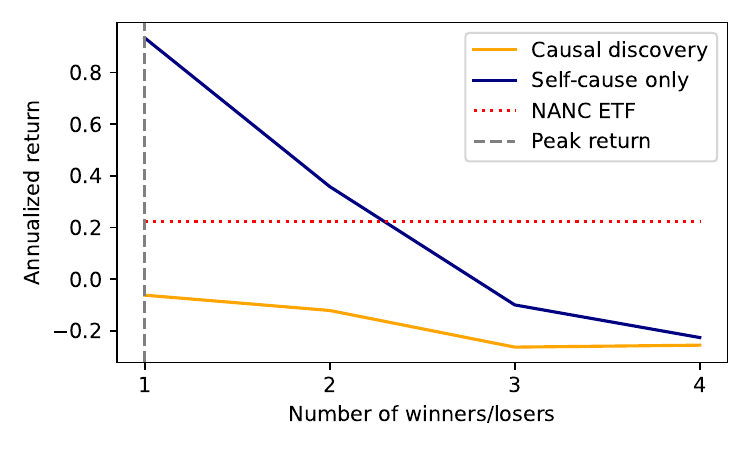}
        \caption{Pelosi, TiMINo, lag=6}
    \end{subfigure}
    \caption{Pelosi portfolio performance using TiMINo}
\end{figure}

\end{document}